\documentclass[3p]{elsarticle}

\usepackage[hypertex]{hyperref}
\usepackage{amsfonts,amssymb}
\usepackage{lineno,hyperref}
\modulolinenumbers[5]

\journal{Physics Letters A}

\begin{document}

\begin{frontmatter}

\title{Fisher information and quantum mechanical models for finance}

\author{V.~A.~Nastasiuk\fnref{myfootnote}}
\address{South Ukrainian National Pedagogical University, Staroportofrankivska Str., 26, Odessa, Ukraine 65020}

\cortext[mycorrespondingauthor]{Corresponding author}
\ead{nasa@i.ua}

\begin{abstract}
The probability distribution function (PDF) for prices on financial markets is derived by extremization of Fisher information. It is shown how on that basis the quantum-like description for financial markets arises and different financial market models are mapped by quantum mechanical ones.
\end{abstract}

\begin{keyword}
Fisher information, Quantum finance
\end{keyword}

\end{frontmatter}

\linenumbers

\section{Introduction}

This work discusses the application of Fisher variation principle \cite{II} to such phenomena as financial markets. Those global formations are considered as complex systems, whose modeling should be carried out by using tools and methodologies of statistical mechanics and theoretical physics.
A complete statistical characterization of different markets (stock, commodities, foreign exchange etc.) includes such important aspect as PDF evaluation. Attempts are ongoing to develop the most satisfactory stochastic model describing the main features encountered in empirical analysis \cite{MA}; the non-Gaussian shape of price returns PDF is one of common themes there.
A second area concerns the development of a theoretical model that is able to encompass the essential features of real financial system which is characterized by such PDF. Financial economics borrows results in statistical physics \cite{VO}, and in addition to statistical-mechanic description, a quantum mechanic representation has also emerged \cite{BA}--\cite{CH}. Unfortunately, the constructing of quantum theory for finances often reduces to direct postulation of Schr\"{o}dinger equation and its subsequent solution under some entry conditions. The purpose of present letter is to show that the quantum mechanical framework can be naturally derived from Fisher information thus providing a comprehensive basis for financial economics.

A very common Fisher-technique replaces the entropy by Fisher information measure in many applications (see e.g. \cite{FR}--\cite{SI}, and list of references therein). But, in contrast to the maximum entropy problem with its solution in a fixed exponential form, the minimizing (extremizing) Fisher information function leads to a second order differential equation of Schr\"{o}dinger type, whose solutions exhibit a variety of mathematical forms.
Shaping the potential function in Schr\"{o}dinger equation we give some 'physics' to rise. Then the specific choice for the solutions has to grasp main features of the empirical statistical distributions.
\section{Extremum Fisher information approach}

It is known that prices on financial markets change randomly remaining near the same in average. Taking a sequence of price changes $x$ (logarithmic returns commonly used in financial analysis) obtained through a fixed time interval and estimating their PDF $p(x)$ we can define the dispersion $\sigma^2=\langle( x-\langle x\rangle)^2\rangle$ of observations, or market \emph{volatility}.  Fisher's information measure \cite{FR}
\begin{equation} \label{eq.1}
I=\int{dx p(x)\left(\frac{d\ln p(x)}{d x}\right)^2}=4\int{dx \left(\frac{d \psi(x)}{d x}\right)^2}
\end{equation}
($\psi(x)\equiv\sqrt{p(x)}$) arises as a lower bound for the dispersion according to the Cramer-Rao inequality \cite{HA}
$$\sigma^2\cdot I\geq 1,$$
 serving thereby as a quality metric of a price decrements predictability.

Suppose, the financial system state can additionally be characterized by mean values \begin{equation}\label{ax} F_k = \int{dx f_k (x)p(x)}, k = 1, ..., M,
\end{equation}
of some $M$ functions $f_k(x)$. The set of measurable values (\ref{ax}) constitutes an available empirical information about system. The statistical moments of very PDF could be taken as $F_k$, if there is no information of other origin.

In this context the relevant PDF extremizes $I$ subject to the prior conditions (\ref{ax}) \cite{II}, \cite{FR}. Normalization entails $\int{dxp(x)}=1$, and the Fisher-based extremization problem adopts the appearance
\begin{equation}\label{var}
\delta_p \left(I[p]-\epsilon\int{dx p(x)}-\sum_{k=1}^M\lambda_k \int{dxf_k(x)p(x)}\right) =0
\end{equation}
with the $(M+1)$ Lagrange multipliers $\epsilon$, and $\lambda_k$ which possible financial meaning is clarified under specific formulation of problem (through a particular choice of set of functions $f_k(x)$).

Variation (\ref{var}) yields now Schr\"{o}dinger equation for the amplitude $\psi(x)$  \cite{II}:
\begin{equation}\label{sch}
-\frac 1 2 \psi''-\frac 1 8 \sum^M_k \lambda_k f_k(x)\psi(x)= \frac{\epsilon}{8}\psi(x),
\end{equation}
where the multiplier $E=\epsilon/8$ plays evidently a role of energy eigenvalue.

Obviously that a choice of the set of functions $f_k(x)$, whose means can be detected, entirely defines the mathematical form of Eq. (\ref{sch}) solutions, and PDFs respectively, and hence the physical model for financial system.
\section{Financial quantum oscillator}
If our prior knowledge about system under investigation is limited to power moments
\begin{equation}\label{*}
F_k=\left< f_k\right>=\left< x^k\right>,
\end{equation}
we can introduce into Schr\"{o}dinger equation (\ref{sch}) the potential function in the manner
\begin{equation}\label{5}
U(x)=-\frac 1 8 \sum_k^M\lambda_k x^k
\end{equation}

It is possible because $U(x)$ belongs to $\mathcal{L}_2$ and thus admits a series of expansion in $x,x^2,x^3$, etc. \cite{FB}.

The number $M$ of terms that should be kept in the potential (\ref{5}), and accordingly, the moments to be fitted, sufficient to approach a goal distribution, depends, in general, on particular distributions. For distribution functions centered at zero the first moment drops out. If one retains in expansion (\ref{5}) the second term only he gets the equation for quantum harmonic oscillator
\begin{equation}\label{*}
-\frac 1 2 \psi''+\frac {\omega^2 x^2} 2 \psi= E\psi
\end{equation}
with the frequency $\omega=\sqrt{|\lambda_2|}/2$ and energy spectrum $$E_n=\omega\left( n+\frac 1 2\right).$$

The empirical price return distributions are typically unimodal. So, among the multiple oscillator's eigenfunctions \cite{FL} (with $H_n(x)$ as Hermite's polynomials)
\begin{equation}\label{*}
\psi_n^0(x)=\sqrt{\frac 1 {2^nn!}\sqrt\frac{\omega}{\pi}}H_n(\sqrt{\omega}x)e^{-1/2\omega x^2},
\end{equation}
one has to choose the ground state solution, $n=0$, which has no nodes. This yields
\begin{equation}\label{*}
p(x)=(\psi^0_0)^2=\sqrt{\frac{\omega}{\pi}}e^{-\omega x^2}.
\end{equation}

It is a Gaussian distribution with dispersion $\sigma^2\equiv\left<x^2\right>=(2\omega)^{-1}=(4E_0)^{-1}$.

The normal PDF is considered by theory of finance often as a first approximation of what is observed in empirical data. In order to take into account the small deviation from normality we could retain more terms in expansion (\ref{5}). With the third and forth moments we come to the problem of unharmonic oscillator:
\begin{equation}\label{*}
-\frac 1 2 \psi''+\left(\frac {\omega^2 x^2} 2+\varepsilon_1(\sqrt{\omega}x)^3+\varepsilon_2(\sqrt{\omega}x)^4\right)\psi= E\psi.
\end{equation}
(Here the re-designations $\lambda_3=-8\varepsilon_1\omega^{3/2}$ and $\lambda_4=-8\varepsilon_2\omega^2$ are performed).

In the frame of quantum mechanical perturbation theory (when the coefficients $\varepsilon_1$ and $\varepsilon_2$ are small enough) the eigenfunction with $n=0$ we are interested in can be written as \cite{LL}
\begin{equation}\label{ex}
\psi_0(x)=\psi_0^0(x)+
\sum_{m\neq 0}\frac{\left<m\mid\varepsilon_1(\sqrt{\omega}x)^3+\varepsilon_2(\sqrt{\omega}x)^4\mid 0\right>}{E_m-E_0}\psi_m^0(x)+...
\end{equation}

As result of standard calculations we obtain in a first approximation the wave function for perturbed ground state
\begin{equation}\label{psi}
\psi_0 (x)=\left(\frac{\omega}{\pi}\right)^{\frac 1 4 }e^{-\frac{\omega x^2} 2}\left[1-\frac{15}{16}\frac{\varepsilon_2}{\omega}-
\frac{2\varepsilon_1}{\sqrt\omega}x+\frac 9 4 \varepsilon_2 x^2+\frac {\sqrt\omega\varepsilon_1} 3 x^3-\frac{\omega\varepsilon_2} 4 x^4\right].
\end{equation}

Correspondingly, the approximated distribution function takes the form
\begin{equation}\label{px}
p(x)=\left(\psi_0 (x)\right)^2=\sqrt\frac{\omega}{\pi}e^{-\omega x^2}C_8(x;\omega,\varepsilon_1,\varepsilon_2)
\end{equation}
where the polynomial of degree eight, $C_8$ is a square of bracketed expression in (\ref{psi}).

The unknown parameters $\omega,\varepsilon_1,\varepsilon_2$ is to be evaluated by fitting (\ref{px}) under empirical curve.
\section{Delta potential and high leptokurtosis}
Although there is no complete agreement in literature with respect to the actual shape of financial returns distributions, some models successfully exploit the idea of power-law tailed PDF (see Refs. in \cite{MA} and \cite{SH}). The fact is that empirical return distributions, while unimodal and approximately symmetric, are typically found to exhibit considerable leptokurtosis, i.e., they are more peaked in the center and have fatter tails than the Gaussian with the same variance. Trying to fit such PDF curve with the function (\ref{px}) does not lead to satisfactory result. So, the information potential $U(x)$ requires modification.
The square potential well
\begin{equation}\label{17}
U(x) =\left\{
    \begin{array}{rcl}
          -\lambda, & \mid x\mid\leq a\\
           0, & \mid x\mid>a
         \end{array}
         \right.
\end{equation}
seems to be a better choice instead of parabolic one. Among solutions of the Schr\"{o}dinger equation with potential (\ref{17}) the even ones have the view \cite{GK}:
\begin{equation}\label{psiwell}
\psi(x)=\left\{\begin{array}{cc}
           A\cos{\sqrt{2(\lambda-|E|)}x}, & |x|\leq a \\
           Be^{-\sqrt{2|E|}|x|}, & |x|>a
         \end{array}
         \right.
\end{equation}
where the eigenvalues $E$ depend on width $a$ and depth $\lambda$ of the well.

The exponential decaying of wave function (\ref{psiwell}) outside the well corresponds to fat tailed PDF, and if the peak of PDF is sharp, the potential well is fine. Upon fineness condition $a^2\lambda\ll 1$, the specific form of wave function middle part ($|x|<a$) loses significance and the only state $$
\psi(x)\approx(2|E|)^{1/4}e^{-\sqrt{2|E|}|x|}
$$
with energy $|E|\approx2\lambda^2a^2$ takes place. In this state the standard deviation $\sigma=(4|E|)^{-1/2}\gg a$, meaning PDF tails fatter than that for Gaussian.

Without going into details of the PDF peak structure one can put $a=0, \lambda\rightarrow\infty$ and transit from (\ref{17}) to delta potential
$$U(x)=-\lambda\delta(x).
$$

This transition means that, according to (\ref{ax}), our prior knowledge about the system regards only to the height of PDF peak:
$$ F=\int dx p(x)\delta(x)=p(0).
$$

Now the unique solution of Schr\"{o}dinger equation is
\begin{equation}\label{*}
\psi(x)=\sqrt\lambda e^{-\lambda \mid x\mid}.
\end{equation}

Corresponding one-parametric PDF,
\begin{equation}\label{la}
p(x)=\lambda e^{-2\lambda \mid x\mid},
\end{equation}
 is the Laplace distribution.

Since $x$ denotes log return, Eq. (\ref{la}) implies that the returns $y=e^x$ of asset price has a power law distribution
$$
p(y)=\frac{\lambda}{\mid y\mid^{2\lambda}}
$$
widely discussed in \cite{SH}.
\section{Summary}
The quantum mechanical structure of financial markets has been expressed using Fisher information. In that context the state of financial system characterized by some PDF is correlated with a coupled state of quantum particle in the space of price decrements (velocities). The potential function of related Schr\"{o}dinger equation originates in prior empirical data which one learns from PDF.
It is shown that deep parabolic potential well corresponds to important model of normally distributed random walk. This model is known else as \emph{Efficient market} in the theory of finance. In opposite the non-Gaussian models of price behavior, describing the PDF's tails decay as exponential, are corresponded by a fine potential well, or even delta-function. It is obvious for further research that PDF specific features could be modeled by variation of potential function form.

The nice feature about our approach is that it results in analytically solvable equations, which invites empirical investigations using widely available financial data.
\section*{References}


\end{document}